\documentclass{PoS}
\usepackage{amsmath}
\title{One-loop finite corrections to seesaw neutrino masses}

\ShortTitle{One-loop finite corrections to seesaw neutrino masses}

\author{Diego ARISTIZABAL\\
       IFPA, Dep. AGO, Universite de Liege, Bat B5 Sart
       Tilman B-4000 Liege 1, Belgium
       \\
       E-mail: \email{daristizabal@ulg.ac.be}}

\abstract{In the standard seesaw model, finite corrections to the
  neutrino mass matrix arise from one-loop self-energy diagrams
  mediated by heavy neutrinos. We discuss the impact that these
  corrections may have on the different entries of the tree-level
  effective neutrino mass matrix, paying special attention to
  their dependence with the seesaw model parameters. We also briefly
  comment on the implications these corrections might have on 
  low-energy neutrino observables.}

\FullConference{XXIst International Europhysics Conference on
  High Energy Physics\\
  21-27 July 2011\\
  Grenoble, Rhones Alpes France}

\begin{document}

\section{Motivation}
\label{sec:mot}
Among the plethora of models for Majorana neutrino masses present in
the literature the standard seesaw \cite{seesaw} is certainly the most
popular mechanism for neutrino mass generation. In this model, by
extending the standard model Lagrangian with three fermionic
electroweak singlets, the five dimensional effective operator $LLHH$
is realized via the exchange of the new states. The smallness of the
light neutrino masses is determined by the suppression induced by the
scale of lepton number violation which is assumed to be large 
${\cal O}(\Lambda_{\text{GUT}})$.

The seesaw parameter space depends upon 18 ``coordinates'': 6 phases
and 12 real parameters. Low-energy data implies -in principle- 9
constraints, provided the absolute light neutrino mass scale and the
Dirac and Majorana CP violating phases are measured. Given the
mismatch between the number of parameters and observables a unique
region consistent with data \cite{Schwetz:2011zk} can not be
fixed. The analysis of the available portions in parameter space is
thus based on scans which in turn rely on parametrizations of the
seesaw. All these parametrizations are based on the tree-level
effective mass matrix, as the one-loop order corrections are assumed
to be negligible. This however might not be the case if the
corrections are finite \cite{Grimus:2002nk,AristizabalSierra:2011mn}.
\section{One-loop finite corrections}
\label{sec:res}
\begin{figure}
  \centering
  \includegraphics[width=7cm,height=2cm]{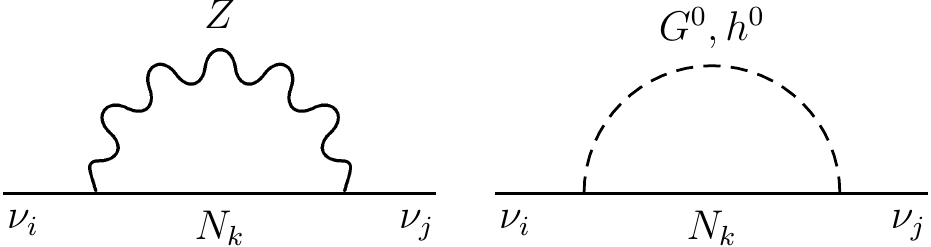}
  \caption{Self-energy diagrams accounting for $\mathbf{\delta M_L}$}
  \label{fig:slef-energy-diagrams}
\end{figure}

The fermionic electroweak singlets $N_R$ induce new interactions
that, in the basis in which the matrix of charged lepton Yukawa
couplings and the singlet mass matrix $\pmb{M_R}$ are diagonal, are
described by the following Lagrangian
\begin{equation}
  \label{eq:seesaw-Lag}
  -{\cal L}=-i\bar N_{R_i}\,\gamma_\mu\partial^\mu N_{R_i}
  + \tilde\phi^\dagger\bar N_{R_i}\lambda_{ij}\ell_{Lj}
  + \frac{1}{2}\bar N_{R_i} C M_{R_i} \bar N_{R}^T
  + \mbox{h.c.}
\end{equation}
Here $\phi^T=(\phi^+ \phi^0)$ is the Higgs electroweak doublet,
$\ell_L$ are the lepton $SU(2)$ doublets, $C$ is the charge
conjugation operator and $\pmb{\lambda}$ is a Yukawa matrix in flavor
space.  In the seesaw limit that is to say, $M_R\gg v$ (with $v\simeq
174$ GeV) the effective neutrino mass matrix can be written according
to
\begin{equation}
  \label{eq:seesaw-formula-tree-level}
\pmb{m_\nu}^{\mbox{\tiny(tree)}}=
-v^2\pmb{\lambda}^T\,\pmb{\hat M_R}^{-1}\,\pmb{\lambda}\,.
\end{equation}
Finite corrections to the above matrix arise from the one-loop
self-energy diagrams shown in figure \ref{fig:slef-energy-diagrams}
and are given by \footnote{For details see
  ref. \cite{Grimus:2002nk} or the appendix in ref. \cite{AristizabalSierra:2011mn}.}
\begin{equation}
   \label{eq:deltaML}
   \pmb{m_\nu}^{\mbox{\tiny(1-loop)}}=
   v^2\pmb{\lambda}^T \pmb{\hat M_R}^{-1}
   \left\{\frac{g^2}{64 \pi^2 M_W^2}
     \left[
       m_h^2\ln\left(\frac{\pmb{\hat M_R}^2}{m^2_h}\right)
       +
       3 M_Z^2\ln\left(\frac{\pmb{\hat M_R}^2}{M^2_Z}\right)
     \right]\right\}\pmb{\lambda}\,.
\end{equation}
Notice that this correction is not suppressed with respect to
the tree-level result by additional factors of
$v\,\pmb{\lambda}\,\pmb{M_R}^{-1}$. Thus, it is expected to be smaller
than the tree-level mass term solely by a factor of order
$(16\pi^2)^{-1}\ln(M_R/M_Z)$, implying it might have sizable effects.
  
\begin{figure}[h]
\begin{center}
\includegraphics[width=12cm,height=4.5cm]{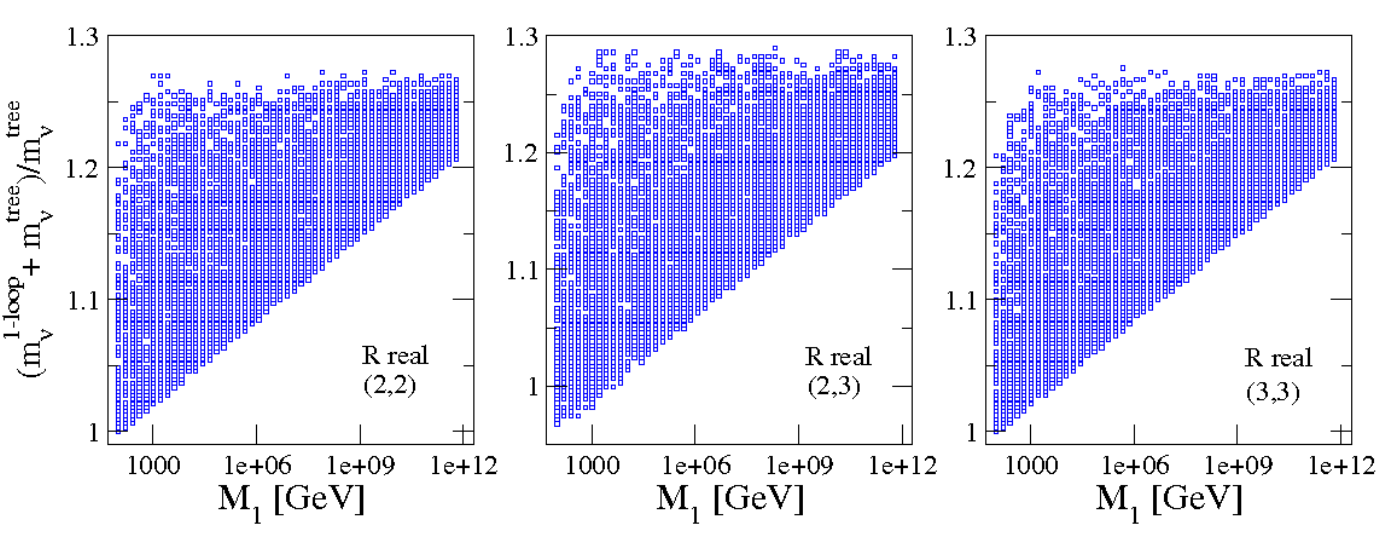}
\caption{Relevance of the finite one-loop correction for the 22, 23
  and 33 elements of the neutrino mass matrix in the case of a real
  $\pmb{R}$ matrix.}
\end{center}
\label{fig:harrayRR-red}
\end{figure}

We evaluate the importance of the correction in (\ref{eq:deltaML}) by
using the Casas-Ibarra parametrization \cite{Casas:2001sr}:
\begin{equation}
  \label{eq:c-i-p}
  \pmb{\lambda}=\frac{\sqrt{\pmb{\hat M_R}}\,\pmb{R}\,\sqrt{\pmb{\hat
        m_\nu}} \,\pmb{U}^\dagger}{v}\,,
\end{equation}
where $\pmb{R}$ is a general complex orthogonal matrix, and scanning
the parameter space assuming a normal hierarchical spectrum and
a real $\pmb{R}$. The results for the 22, 23 and 33 elements of the mass
matrix are displayed in figure \ref{fig:harrayRR-red}.

An analysis of how the finite one-loop corrections may affect for
example the neutrino mixing angles can be carried out by assuming a
well motivated mixing scheme as an input. This has been done in ref.
\cite{AristizabalSierra:2011mn} (including also a study of the
neutrino mass spectrum) where it has been shown that even in
conservative scenarios the effects can be sizable. In conclusion, due
to their relevance we argue these corrections must be taken into
account in the study of the seesaw parameter space.


\begin{thebibliography}{99}
\bibitem{seesaw} 
  P. Minkowski, {\it Phys. Lett.} B {\bf 67} 421
  (1977); T.  Yanagida, in {\it Proc. of Workshop on Unified Theory
    and Baryon number in the Universe}, eds. O. Sawada and
  A. Sugamoto, KEK, Tsukuba, (1979) p.95; M. Gell-Mann, P. Ramond and
  R. Slansky, in {\it Supergravity}, eds P.  van Niewenhuizen and
  D. Z. Freedman (North Holland, Amsterdam 1980) p.315; P.  Ramond,
  {\it Sanibel talk}, retroprinted as hep-ph/9809459; S. L. Glashow,
  in{\it Quarks and Leptons}, Carg\`ese lectures, eds M. L\'evy,
  (Plenum, 1980, New York) p. 707; R. N. Mohapatra and
  G. Senjanovi\'c, {\it Phys. Rev.  Lett.} {\bf 44}, 912 (1980);
  J.~Schechter and J.~W.~F.~Valle,
  Phys.\ Rev.\  D {\bf 22} (1980) 2227;
  Phys.\ Rev.\  D {\bf 25} (1982) 774.

\bibitem{Schwetz:2011zk} 
  T.~Schwetz, M.~Tortola, J.~W.~F.~Valle,
  New J.\ Phys.\  {\bf 13}, 109401 (2011).
  [arXiv:1108.1376 [hep-ph]].

\bibitem{Grimus:2002nk}
  W.~Grimus, L.~Lavoura,
  Phys.\ Lett.\  {\bf B546}, 86-95 (2002).
  [hep-ph/0207229].

\bibitem{AristizabalSierra:2011mn}
  D.~Aristizabal Sierra, C.~E.~Yaguna,
  JHEP {\bf 1108}, 013 (2011).
  [arXiv:1106.3587 [hep-ph]].

\bibitem{Casas:2001sr}
  J.~A.~Casas, A.~Ibarra,
  Nucl.\ Phys.\  {\bf B618}, 171-204 (2001).
  [hep-ph/0103065].
\end{thebibliography}
\end{document}